\documentclass[12pt]{article}
\makeatletter
\def\section{\@startsection {section}{1}{\z@}{-3.5ex plus -1ex minus
 -.2ex}{2.3ex plus .2ex}{\large\bf}}
\def\subsection{\@startsection{subsection}{2}{\z@}{-3.25ex plus -1ex 
minus -.2ex}{1.5ex plus .2ex}{\normalsize\bf}}

\newcommand{\be}{\begin{equation}}
\newcommand{\ee}{\end{equation}}

\begin{document}
\begin{titlepage}
\rightline{NORDITA-2000/123 HE}
\vskip 1.8cm
\centerline{\Large \bf STABLE NON-BPS D-BRANES} 
\vskip 0.4cm
\centerline{\Large \bf  AND THEIR CLASSICAL DESCRIPTION
\footnote{Work partially supported by the European Commission RTN
programme HPRN-CT-2000-00131 and by MURST.}}
\vskip 1.4cm 
\centerline{\large \bf M. Bertolini $^a$ and A. Lerda $^{b,c}$}
\vskip .8cm \centerline{\sl $^a$ NORDITA, Blegdamsvej 17, DK-2100
Copenhagen \O, Denmark}
\vskip .4cm \centerline{\sl $^b$ Dipartimento di Scienze e Tecnologie
Avanzate} \centerline{\sl Universit\`a del Piemonte Orientale, I-15100
Alessandria, Italy}
\vskip .4cm \centerline{\sl $^c$ Dipartimento di  Fisica Teorica,
Universit\`a di Torino} \centerline{\sl and I.N.F.N., Sezione di
Torino,  Via P. Giuria 1, I-10125 Torino, Italy}
\vskip 1.8cm
\begin{abstract}
We review how to describe the stable non-BPS D-branes of type II string theory
from a classical perspective, and discuss the properties
of the space-time geometry associated to these configurations.
This is relevant in order to see whether and how the gauge/gravity
correspondence can be formulated in non-conformal and non-supersymmetric
settings.
\end{abstract}
\end{titlepage}
\renewcommand{\thefootnote}{\arabic{footnote}}
\setcounter{footnote}{0} \setcounter{page}{1} 

\tableofcontents  
\vskip 1.5cm
\section{Introduction}
D-branes have played a crucial role in almost all recent developments
of string theory. This is mostly due to the fact that they admit a
simple and explicit description in terms of open strings with Dirichlet 
boundary conditions; moreover, being BPS saturated
objects, they allow to obtain relevant information about 
non-perturbative aspects of string theory.
In the last couple of years, however, after a remarkable series of 
papers by A. Sen \cite{Sen1,Sen2,Sen4,Sen5,Sen6},
a lot of attention has been devoted also to the study of non-BPS
D-branes (for reviews see Ref.s~\cite{Senr,leru,schw1,gablec}). 
There are at least two main motivations which justify this
kind of studies. The first one is that non-BPS D-branes 
are useful in testing various string dualities
beyond the BPS level in order to put them on a stronger and
more concrete basis.
The second motivation, which is even more ambitious,
has to do with possible extensions of the Maldacena 
gauge/gravity correspondence to
non-conformal and less supersymmetric or even non-supersymmetric
settings. To this purpose, a preliminary but necessary step 
is represented by the study of the
classical geometry generated by the non-BPS branes. 
In this paper we review the recent results
which have been achieved in  this direction, mainly referring to
Ref.~\cite{noi}. 
\section{Stable non-BPS D-branes}
It is by now well-known that type II string theories in ten dimensions
possess non-BPS D$p$-branes, with $p$ odd for type IIA and $p$ even for 
type IIB. These non-BPS D-branes do not preserve any supersymmetry
and are unstable 
due to the existence of open string tachyons on
their world-volumes. Indeed, they
can decay either into lower dimensional
stable objects ({\it e.g.} BPS D-branes) or into the closed string vacuum
depending on which configuration the tachyons acquire.
The instability of these non-BPS D-branes
clearly puts severe limitations on their use in finding
a non-supersymmetric extension of the gauge/gravity correspondence, 
since a stable geometric background seems to be a crucial 
and necessary ingredient to this purpose.
As a matter of fact, for the non-BPS D-branes of type II in flat
space one cannot disentangle their non-supersymmetric nature from their
instability; therefore, despite their intrinsic interest, 
it is not immediately obvious to say whether or not they can yield
a classical geometry and so
their use and applications remain doubtful.

Luckily, however, there exist orientifolds and
orbifolds of type II theories which admit {\it stable} non-BPS
D-branes! This happens whenever the orientifold/orbifold projection
is able to remove any tachyonic state from the open string spectrum
on the brane world-volume, so that the resulting configuration 
is stable even if it is not supersymmetric. 
For example, in the type I theory, which is
an orientifold of  the  type IIB string with respect to the world-sheet
parity, it turns out that the D-instanton and the D-particle
are stable non-BPS branes \cite{Sen4,gallot}. On the other hand, 
the existence of such D-branes is required by the type~I/SO(32) heterotic
string duality, and indeed their properties and interactions are 
in agreement with this duality \cite{gallot1}. 
Another well-studied model which possesses
stable non-BPS D-branes is the type IIA string on the
orbifold $T_4/ {\cal I}_4$, where $T_4$ is a 4-torus   
and ${\cal I}_4$ is the parity along its directions.
This theory is a singular limit of the type IIA string compactified on
a K3 manifold and is related by a non-perturbative duality to
the heterotic string compactified on $T_4$. From such a duality
it is possible to infer that the IIA orbifold model possesses
several non-BPS but stable D-branes, like for
example the non-BPS D-string \cite{Sen4,Gab1}.
If one performs a T-duality along one of the four compact
directions, one obtains the type IIB theory on the orbifold
$T_4/(-1)^{F_L}{\cal I}_4$, where  $(-1)^{F_L}$ is 
the contribution to the space-time fermion
number from the left-moving closed string sector.
This is the theory which we consider in the remaining part of this paper.

Let us take for instance 
a non-BPS D$p$-brane of type IIB ({\it i.e.} $p$ even)
whose world-volume is completely transverse
with respect to the orbifold directions; 
then it can be shown that the 
tachyonic NS ground state of the open string living on the 
brane world-volume, is globally {\it odd} under the orbifold
parity and is projected out. The lighest states of the
would-be tachyon field which can survive the orbifold projection
are therefore the windings modes of level one. In fact, these modes are odd
under $(-1)^{F_L}$ (because for open string states 
this operator has the  same action as the GSO parity $(-1)^F$), 
and hence combinations of level-one winding states which are 
odd under ${\cal I}_4$ are
globally even and survive the orbifold projection. 
More precisely,
there are four such states that correspond to the four possible
parity-odd combinations along the four compact directions,
which we take to be $\{x^6,x^7,x^8,x^9\}$, and that are defined as follows
\begin{equation}
|\chi_a\rangle \equiv |k^\alpha ; 1_a \rangle - |k^\alpha ; - 1_a
\rangle\qquad (a=6,...,9) 
\label{state}
\end{equation}
where $k^\alpha$ is the momentum along the flat directions and $\pm 1_a$
represents positive or negative winding of level one along the
$a$-th direction. The mass of these states is given by 
\begin{equation}
 M_a^2 =
\frac{R_a^2}{\alpha'^2} - \frac{1}{2\alpha'} 
\end{equation}
where $R_a$ is the compactification radius of the $a$-th coordinate. Thus
we easily see that for
\begin{equation}
R_a \geq \sqrt{\alpha'/2} \equiv R_{crit}
\label{bound}
\end{equation}
the lightest state of the would-be tachyon field that is selected by
the orbifold parity has a non-negative
mass squared. In this case, no tachyonic excitation 
is present on the world-volume
and hence the non-BPS
D$p$-brane is truly stable.
A similar analisys holds in various $T$-dual
situations. Clearly, when  the $T$-duality is performed along an
orbifolded direction, there is an exchange between  windings and
momenta, and the above bound gets exchanged to  $R_a \leq \sqrt{2
\alpha'}$  for the given direction. 

The stability of these non-BPS states can be understood also from a
closed  string point of view. The reason why type II non-BPS D-branes
in ten dimensions are unstable is essentially due to the fact that
they are neutral objects which are not charged under any R-R
field: therefore, their tension is not balanced by any charge ``repulsion''
which could stabilize them, as it happens for BPS D-branes. Upon
orbifold compactification, while these states remain neutral under R-R
{\it untwisted} fields, they actually  acquire a charge under {\it
twisted} R-R ones, and when the bound of Eq. (\ref{bound}) is
satisfied  their charge is sufficient to balance their tension and
stabilizes them.

Our next task is to use these non-BPS branes as elementary sources
to produce a classical
geometry. To this aim, the stability which we have discussed
above, is of course a necessary but not sufficient requirement.
In fact, one needs also to have 
big charges, in string units, in order to ensure 
small curvatures and hence the possibility of
neglecting higher derivative terms in the low energy effective theory.
In other words one should be able to 
construct a bound state made of a large number
$N$ of microscopic constituents, and this is quite difficult for
non-BPS D-branes. However, a remarkable property holds at the border
of the stability region defined by Eq.(\ref{bound}): in fact, when the
compactification radii are tuned to the critical value $R_{crit}$, 
the world-volume states of Eq.(\ref{state}) become massless and an
accidental bose-fermi degeneracy occurs on the non-BPS D-brane \cite{gabsen}. 
As a consequence, the one-loop open string partition function vanishes. 
From the closed string point of view, this property is expressed
as the vanishing of the tree level exchange amplitude between 
two boundary states
\begin{equation}
\label{0force}
Z =\langle Dp|\,{\cal P}\,|Dp\rangle = 0
\end{equation}
where ${\cal P}$ is the closed string propagator and $|Dp\rangle$ is
the  boundary state representing the D$p$-brane as a
source of closed string states (for a review on the boundary
state formalism and its applications, see  Ref.\cite{anto}). 
This means that at critical radii the force between two 
non-BPS D-branes vanishes, at one
loop. In this case, then, the non-BPS D-branes enjoy two properties:
they are stable and  a no-force condition holds. This clearly
opens-up the possibility of constructing macroscopic bound states 
with these microscopic objects and obtain a classical
geometry associated to them.  More precisely, it is natural to
consider a superposition of a large
number of these stable non-BPS D-branes and simply take the na\"{\i}ve sum
of $N$ ``single particle'' boundary states, that is 
\begin{equation}
\label{NB}
|Dp,N\rangle \equiv N \, |Dp\rangle \quad 
\end{equation}
for large $N$. This is the working hypothesis which we are 
going to test in the next section.
\section{The non-BPS D-particle solution}
In this section we consider as a concrete example 
the non-BPS D-particle
of type IIB on $T_4/(-1)^{F_L}{\cal I}_4$ and look for the corresponding 
classical geometry. However, all relevant 
features of our analysis are to a large extent
independent of the specific example considered and are shared by 
essentially all stable non-BPS D-branes arising in (four dimensional)
orbifold  compactifications of type II string theories. 

The classical description of the non-BPS D-particle can be
derived by solving the equations of motion that describe the dynamics of
the supergravity fields emitted by the D-particle itself. 
In this case, such equations are those of 
the non-chiral ${\cal N}=(1,1)$ supergravity in six dimensions.
However, not all fields of this theory 
are coupled to the D-particle, and to see 
which ones should be considered, we use the 
boundary state formalism~\cite{anto}. In this framework, 
the non-BPS D-particle is described
by a generalized coherent state
$|D0\rangle$ which has a NS-NS
untwisted component and a R-R twisted one. 
Its structure is thus of  the form
\begin{equation}
\label{D0}
|D0\rangle = |D0\rangle_{{\rm NS-NS},U} + |D0\rangle_{{\rm R-R},T_I}~~,
\end{equation}
where the index $I=1,...,16$ in the twisted part indicates on which
orbifold plane the D-brane is placed.
By computing the overlaps of
$|D0\rangle$ with the  perturbative closed string states, one gets two
essential information: {\it i)} which are the bulk fields emitted
by the brane; {\it ii)} the asymptotic behavior at large distances
of these fields \cite{class}.
From this information,
the full action describing the dynamics of  the
fields coupling to the non-BPS D-particle can be easily reconstructed
and it turns out to be given by the sum of a bulk
action, which is a consistent truncation of the ${\cal
N}=(1,1)$ supergravity Lagrangian, and a world-volume action, which
acts as a source term. 

The explicit form of the coherent state $|D0\rangle$ in Eq.(\ref{D0})
and their overlaps with perturbative closed string states have been
studied for example in Ref.s \cite{eyras,noi}. 
The result of this analysis is
that the six dimensional fields that couple to the 
non-BPS D-particle are: the graviton
$G_{\mu\nu}$,  five scalars, {\it i.e.} $\varphi$ (related to the ten
dimensional dilaton) and $\eta_a$ ($a=1,...,4$) (related to the internal
components of the ten dimensional metric), and finally one (twisted)
vector field $A_\mu$. The action describing the dynamics of these
fields at critical compactification radii is 
\begin{eqnarray}
\label{sum1}
S&=&\frac{1}{2\kappa_{orb}^2}\int d^6x \,\sqrt{-\det G}\left[{\cal
R}(G) - \partial_\mu\varphi\,\partial^\mu\varphi - \partial_\mu\eta_a
\,\partial^\mu\eta_a - \frac{1}{4}\,  {\rm
e}^{\varphi}\,F_{\mu\nu}F^{\mu\nu}\right]  \nonumber \\ &&-\,M  \int
d\tau \,{\rm e}^{-\frac 12 \varphi - \frac 12 \sum_a \eta_a}
\sqrt{-G_{00}} \,+ \,M\int \,A_{(1)} ~~,
\end{eqnarray}
where the first line refers to the bulk contribution, the second line
to the boundary contribution, and $\kappa_{orb}^2=32\pi^3\alpha'^2g^2$,
$g$ being the closed string coupling constant. 
Some comments are in order at this point:

{\it i)} The boundary action is not the most general one, since
all world-volume fields have been set to zero. However,
this is the simplest possible
choice which is consistent with our working hypothesis encoded in 
Eq.(\ref{NB}).

{\it ii)} The constant $M$ in front of the DBI term is equal to $N
M_0$  where $M_0$ is the mass of a single non-BPS D-particle ($M_0\sim
(\sqrt{\alpha'} g)^{-1}$).
This is again a consequence of Eq.(\ref{NB}). 

{\it iii)}  Despite the absence of a ``mass=charge'' relation for
a non-BPS brane, the same constant $M$ appears in front of the
DBI and WZ terms of Eq.(\ref{sum1}). This fact occurs only
at the critical radii and can be understood as follows. 
The relative normalization of the gravitational and gauge terms
of the world-volume action can always be adjusted with suitable rescalings of
the various fields. In particular one can always make the coefficient
of the WZ term become equal to the coefficient of the DBI term
by rescaling the R-R potential; but if one does this, in general the 
corresponding kinetic term in the bulk action acquires a non-canonical 
normalization. However, at  critical radii the rescaling which yields 
the same number in front of the DBI and
WZ terms, is precisely the same 
which also gives canonically normalized fields in
the bulk action.  Hence, in this sense one can speak of a ``mass=charge''
relation even for a non-BPS configuration. This fact can be regarded as
the supergravity counterpart of the bose-fermi
degeneracy occuring on the brane's world-volume at critical radii. 

We have now all ingredients to see whether a classical description
for stable non-BPS D-branes is possible. Although the
field equations derived from the action (\ref{sum1})
describe a non-BPS configuration, quite surprisingly it is possible
to explicitly write the solution in a simple
and closed form. In fact, assuming a static and spherically symmetric 
{\it Ansatz}, flat asymptotic space geometry,
and vanishing asymptotic values for the scalar and gauge fields, 
the non-BPS D-particle solution turns out to be~\cite{noi}
\begin{eqnarray}
\varphi&=&\frac{1}{4}\,\ln\left[1+\sin\left(\frac{Q}{r^3}\right)\right]
\label{dil3} \\
\eta_a&=&\frac{1}{4}\,\frac{Q}{r^3} 
\label{scala3} \\
A_0 &=& -1 +
\frac{\cos\left(\frac{Q}{r^3}\right)}{1+\sin\left(\frac{Q}{r^3}\right)}
\label{vec3} \\
G_{00}&=& - \left[1+\sin\left(\frac{Q}{r^3}\right)\right]^{-3/4} 
\label{met30} \\
G_{ij}&=& \delta_{ij}
\left[1+\sin\left(\frac{Q}{r^3}\right)\right]^{1/4}
\label{met3}
\end{eqnarray}
where $Q \equiv 2 M \kappa^2/3\Omega_4 \sim N g \alpha'^{3/2}$, and the indices
$i,j$ label the transverse directions.

This solution is well defined only for $r > Q^{1/3} \sim (g
N)^{1/3}\sqrt{\alpha'}$ and exhibits a naked singularity at
\begin{equation}
r_0 = \left(\frac{2Q}{3\pi}\right)^{1/3}\sim \; 0.6 \; Q^{1/3}
\end{equation}
where the scalar curvature diverges. Within a string theory
perspective, naked singularities are allowed if they occur at distances $\leq
\sqrt{\alpha'}$ where the supergravity approximation is no longer
expected to be valid. From the definition of
$Q$, we see that this is what happens for a single D-particle ($N=1$):
the singularity indeed shows-up at a substringy scale and thus the solution
(\ref{dil3})-(\ref{met3}) is acceptable. 
Moreover, at this scale, the low energy
theory becomes effectively ten-dimensional since the compact dimensions 
are of the same order of $r_0$ (see Eq.(\ref{bound})),
so one is not only neglecting  stringy corrections but the full
Kaluza-Klein tower of massive states. The validity of the low-energy 
solution for a single non-BPS D-particle is consistent with the fact 
that such state, from a string theory perspective, is stable and expected 
to exist. However, as we mentioned earlier, 
a reliable and smooth classical geometry should correspond 
to big charges, {\it i.e.} $N\gg 1$. In this case
the singularity shows-up in a region where the supergravity
approximation is expected to be valid, since now $r_0 \sim Q^{1/3}
\sim (g N)^{1/3} \sqrt{\alpha'}$ and $g N \gg 1$, and
thus the multi D-particle solution should be rejected. 
One can say that while its  
long-distance behavior is consistent
and well defined, the supergravity description is probably
missing some hidden phenomenon occurring at distances $r \sim
r_0$. 

There is another feature of our solution that shows that something
is missing. In fact, the no-force condition, which holds at
one-loop, is broken at higher loops. To understand this point one
should keep in mind that Eq.(\ref{0force}) 
corresponds to a one-loop computation in the open
string coupling constant (or, better, in the 'tHooft coupling $\lambda \sim N
g$ which is proportional to $Q$). The supergravity approximation instead is
valid in a very different regime since it is perturbative in $\alpha'$
but {\it exact} in $\lambda$, and it captures the
low-energy contribution at {\it all} loops in the open string
coupling. Thus to compare with Eq.(\ref{0force}), we must
take the limit $Q\sim\lambda \rightarrow 0$. If we expand our
solution (\ref{dil3})-(\ref{met3}) to first oder in $Q$
and substitute the resulting expressions into the boundary action, we do
find a vanishing force. Indeed, subtracting the vacuum energy, we have
\begin{eqnarray}
S_{\rm boundary}&=& - M \int d\tau \,\,{\rm e}^{-\frac 12 \varphi -
\frac 12 \sum_a \eta_a } \,\, \sqrt{-G_{00} } + M \int d\tau \,A_0
\nonumber \\ &\sim&  - M \int d\tau
\,\,\frac{Q}{r^3}\,\left(-\frac{1}{8} - \frac{1}{2} - \frac{3}{8}  +
1\right) = 0 ~~,
\label{bound3}
\end{eqnarray}
as expected. A similar calculation  however shows that the no-force
condition is {\it not} satisfied at next-to-leading orders, giving
evidence that the  result of Ref.~\cite{gabsen} is spoiled at higher
loops. The same kind of conclusion  has been derived from a different
perspective, namely from a world-volume analysis, in Ref. \cite{lamb1}. 
As we have mentioned earlier, 
the world-volume Lagrangian describes the dynamics  of a set of fields whose
spectrum becomes supersymmetric at the critical radii.  
Nevertheless, the Lagrangian itself is not
supersymmetric and the interactions break the bose-fermi degeneracy, 
thus spoiling the one-loop result of
Eq.(\ref{0force}). In fact, by taking into account
higher loop contributions one can show that non-BPS branes described by a
world-volume action like that of Eq.(\ref{sum1}) repel each other
\cite{lamb2}.
\section{Discussion and conclusions}
We have shown that is possible to provide a supergravity  solution for
stable non-BPS D-branes of the type II string on a K3 orbifold in the
simplest setting suggested by  the one-loop result of
Eq.(\ref{0force}) and described by Eq.(\ref{NB}). However, the
solution represented by Eq.s (\ref{dil3})-(\ref{met3}) has two crucial
drawbacks: it is singular, and the no-force condition does not to hold
when one takes into account higher loop contributions in the open
string coupling. These facts indicate that one should go beyond the
present analisys, and modify the working hypotheses.

Actually, the two drawbacks are not on the same footing and are not
necessarily related to each other.  As mentioned above, to have a
reliable classical solution within a string context, it is necessary
to construct a macroscopic bound state of single  microscopic
constituents. Since for the non-BPS D-particles the no-force
condition does not hold beyond one loop, the appropriate macroscopic
bound state must differ from the n\"aive superposition defined by
Eq.(\ref{NB}). To find the correct bound state is therefore the first
problem one has to solve. One possibility would be to consider some
non-trivial  world-volume dynamics, which can radically change the
structure of the D-brane and its coupling to the bulk fields. This
would lead to an effective action sensibly different  from that of
Eq.(\ref{sum1}), and to new field equations.  In Ref. \cite{lamb2}, by
taking into account string loop corrections to the effective potential
of the world-volume theory, it has been found out that for the stable
non-BPS  D-particle the vacuum  at $\chi_a = 0$ (which is the one
considered here) is a {\it local} minimum of the effective potential
$V$ while the absolute minimum occurs for (some) $\chi_a \not = 0$,
where the non-BPS D-particles attract each other. This fact indicates
that within this setting it  could be possible to construct a
macroscopic bound state corresponding  to $N\gg 1$ non-BPS D-branes
and study the corresponding classical  geometry. Notice that the above
result would not only translate into a change of the boundary action
but also of the bulk action. In fact, at  $\chi_a \not = 0$ the
non-BPS states are in the $D/\bar D$ fractional-brane  phase and
therefore couple also to R-R untwisted and NS-NS twisted
fields. Therefore, the corresponding consistent truncation of the full
${\cal N}=(1,1)$ supergravity action will be different from the one
leading to the action of Eq.(\ref{sum1}). Furthermore, one might
wonder whether following this strategy, it is possible to
automatically  solve also the second drawback affecting the solution
(\ref{dil3})-(\ref{met3}), namely its singular behavior.  This
possibility of course deserves and requires further investigation.

In the previous section we have mentioned the fact that the
supergravity theory could miss some new physics occurring at distances
$r \sim r_0$.  A closer look at our singular solution shows that the
singularity is actually a {\it repulson} \cite{kal}: the gravitational
force vanishes at some distance $r_e > r_0$ and in the region $r_0 < r
< r_e $ the gravitational force is repulsive!  It is by now
well-understood that this property is shared by many other brane
solutions which are dual to non-conformal gauge theories,
independently of the amount of preserved supersymmetries
\cite{KLEBA2,MILANO,ZAMORA, FREEDMAN,TATAR,john,noi2,polch,musto}. In
Ref.\cite{john} a very  interesting stringy mechanism has been
proposed to excise the singularity and yield a regular solution:
independently of the specific details, it turns out that the
appropriate source is not point-like in the transverse space, as one
could have expected,  but rather it is smeared-out on an hypershell
called {\it enhan\c{c}on} locus. In other words, the constituent
branes are forced to cover uniformly an hypersphere rather than pile
up in a single point.  In the simplest cases the {\it enhan\c{c}on} is
located  precisely at $r=r_e>r_0$ and thus the expanded source
surrounds the  singularity at $r=r_0$. While the supergravity solution
remains unchanged for $r>r_e$, it gets drastically modified in the
interior and eventually the singularity is removed.  This mechanism
has been proved to work in various  cases, like for example in the
supergravity solutions which are dual to renormalization group flows
from ${\cal N}=4$ to ${\cal N}=2$ or ${\cal N}=1$ super Yang-Mills
gauge theories \cite{pet1}, or in fractional branes of type II
orbifolds \cite{noi2,polch,musto} (from which the non-BPS D-branes can
be obtained as bound states). All these examples are supersymmetric,
and  it would be very interesting to see whether the {\it
enhan\c{c}on}  mechanism can work also for non-BPS configurations
\footnote{A non-supersymmetric analogue of the enhan\c{c}on mechanism
has been recently discussed in \cite{minic}.}. For this, however, it is
mandatory to have a better understanding of the low-energy dynamics on
the world-volume of non-BPS D-branes, also at loop level.  Possibly,
the results of \cite{lamb2} could help also in this respect: as
already noticed, the true minimum of the stable non-BPS system is
claimed to occur at $\chi_a\not = 0$ which corresponds to the
$D/\bar{D}$ fractional-brane phase. In this phase the structure of the
boundary state is more easely accessible and the relation to the  pure
supersymmetric fractional brane case is manifest.

When studying any kind of gauge/gravity correspondences, it is
necessary to have some control on either side of the duality. This is
the reason why one has to understand which are the microscopic objects
giving rise to a given supergravity solution. Nevertheless, novel
supergravity solutions are also interesting in their own right.  Let
us then end with a pure supergravity remark. The supergravity field
equations one has to solve starting from a bulk action like that of
Eq.(\ref{sum1}), are second order differential equations and their
general solution depends on a certain number of integration
constants. These are uniquely fixed by imposing some general physical
requirements ({\it e.g.} asymptotic flat geometry,  spherical
symmetry, etc.) and by the specific form of the world-volume action,
which acts as a source term. Allowing for different types of bound
states therefore corresponds to repeat the above analysis  and relax
those constraints imposed by the boundary action.  Under these
assumptions,  one can see that the corresponding general solution
depends  on some free parameters related to the mass, the charge and
the dilaton and scalar couplings of the given source.  Moreover, the
periodic functions in Eq.s (\ref{dil3})-(\ref{met3}) get replaced by
hyperbolic ones and the general solution reads
\begin{eqnarray}
\label{solgen} 
e^{\eta_a} &=& \left(\frac{f_{-}(r)}{f_{+}(r)}\right)^{\delta}  \\
\label{solgen1} 
e^{2\varphi} &=& \left(\frac{\cosh X(r) + \gamma \sinh X(r)}{\cosh
\alpha + \gamma \sinh \alpha} \right)
\left(\frac{f_{-}(r)}{f_{+}(r)}\right)^{\frac{3}{4}\epsilon} \\
\label{solgen2}  
A_0 &=&  \sqrt{2(\gamma^2 - 1)} \left(\frac{\sinh X(r) \left(\cosh
\alpha + \gamma \sinh \alpha\right) }{\cosh X(r) + \gamma \sinh X(r)}
- \sinh \alpha \right) \\
\label{solgen3} 
G_{00} &=& - \left(\frac{\cosh X(r) + \gamma \sinh X(r)} {\cosh \alpha
+ \gamma \sinh \alpha}\right)^{-\frac{3}{2}}
\left(\frac{f_{-}(r)}{f_{+}(r)}\right)^{\frac{3}{8}\epsilon} \\
\label{solgen4} 
G_{ij} &=& \delta_{ij} \left(\frac{\cosh X(r) + \gamma \sinh
X(r)}{\cosh \alpha + \gamma \sinh \alpha}\right)^{\frac{1}{2}}
\left(f_{-}(r)\right)^{\frac{2}{3} - \frac{1}{8}\epsilon}
\left(f_{+}(r)\right)^{\frac{2}{3} + \frac{1}{8}\epsilon}
\end{eqnarray}
where
\begin{equation}
\label{f300} 
f_{\pm}(r) = 1 \pm x\, \frac{Q}{r^3} ~~~,~~~ X(r) = \alpha + \beta \ln
 \left(\frac{f_{-}(r)}{f_{+}(r)} \right)
\end{equation} 
and $\alpha$, $\beta$, $\gamma$, $\delta$, $\epsilon$ and $x$ are
integration constants which obey
\begin{equation}  
\epsilon = \pm \frac{4}{3}\sqrt{4 - 3 \beta^2 - 12 \delta^2} ~~.
\label{gensol}
\end{equation}
If in the general solution (\ref{solgen})-(\ref{solgen4}) one imposes
the behavior dictated by the boundary state (\ref{NB}) or equivalently
by the boundary  action of Eq.(\ref{sum1}), then one obtains the
singular solution of Eq.s (\ref{dil3})-(\ref{met3}), but there are
other choices of the parameters which instead lead to singularity-free
solutions (for details see Ref.\cite{noi}). It would be very
interesting to investigate  further these regular solutions and
eventually find which are their  microscopic stringy constituents, if
any.

The highly non-trivial role that stable non-BPS D-branes can play to
obtain a non-conformal and non-supersymmetric extension of  the
AdS/CFT correspondence makes all this kind of investigations quite
challenging.

\vskip 1.5cm
\noindent {\large \bf Acknowledgements}
\smallskip

\noindent 
We thank the co-authors of \cite{noi}, with whom we share all the
results presented in this paper, for the  stimulating and friendly
collaboration. This work is partially supported by the European
Commission RTN programme HPRN-CT-2000-00131. M.B. acknowledges support
by INFN and interesting discussions with M. Gaberdiel, N. Lambert and
I. Sachs.

\end{document}